\documentclass[manuscript]{aastex}
\usepackage{enumerate}
\shorttitle{Observation of a 3D magnetic null point}
\shortauthors{Romano et al.}

\begin{document}


\title{Observation of a 3D magnetic null point}


\author{P. Romano\altaffilmark{1}, M. Falco\altaffilmark{1,2}, S.L. Guglielmino\altaffilmark{3}, and M. Murabito\altaffilmark{3}}%\affil{Astronomy Department, University of California,
%    Berkeley, CA 94720}

%\author{C. D. Biemesderfer\altaffilmark{4,5}}
%\affil{National Optical Astronomy Observatories, Tucson, AZ 85719}
\email{prom@oact.inaf.it}

%\and

%\author{R. J. Hanisch\altaffilmark{5}}
%\affil{Space Telescope Science Institute, Baltimore, MD 21218}


\altaffiltext{1}{INAF - Osservatorio Astrofisico di Catania,
              Via S. Sofia 78, 95123 Catania, Italy.}
\altaffiltext{2}{INAF - Osservatorio Astronomico di Roma, 
							Via Frascati 33, I-00040 Monte Porzio Catone, Italy.}							
\altaffiltext{3}{Dipartimento di Fisica e Astronomia - Sezione Astrofisica, Universit\`{a} di Catania,
			 Via S. Sofia 78, 95123 Catania, Italy}
		 			 
%\altaffiltext{3}{present address: Center for Astrophysics,
%    60 Garden Street, Cambridge, MA 02138}
%\altaffiltext{4}{Visiting Programmer, Space Telescope Science Institute}
%\altaffiltext{5}{Patron, Alonso's Bar and Grill}


\begin{abstract}
We describe high resolution observations of a GOES B-class flare characterized by a circular ribbon at chromospheric level, corresponding to the network at photospheric level. We interpret the flare as a consequence of a magnetic reconnection event occurred at a three-dimensional (3D) coronal null point located above the supergranular cell. The potential field extrapolation of the photospheric magnetic field indicates that the circular chromospheric ribbon is cospatial with the fan footpoints, while the ribbons of the inner and outer spines look like compact kernels. We found new interesting observational aspects that need to be explained by models: 1) a loop corresponding to the outer spine became brighter a few minutes before the onset of the flare; 2) the circular ribbon was formed by several adjacent compact kernels characterized by a size of 1\arcsec-2\arcsec; 3) the kernels with stronger intensity emission were located at the outer footpoint of the darker filaments departing radially from the center of the supergranular cell; 4) these kernels start to brighten sequentially in clockwise direction; 5) the site of the 3D null point and the shape of the outer spine were detected by RHESSI in the low energy channel between 6.0 and 12.0 keV. Taking into account all these features and the length scales of the magnetic systems involved by the event we argued that the low intensity of the flare may be ascribed to the low amount of magnetic flux and to its symmetric configuration.

\end{abstract}


\keywords{Sun: photosphere --- Sun: chromosphere --- Sun: flares --- Sun: magnetic fields}


\section{Introduction}
Flares occur on the Sun and involve different layers of the solar atmosphere. The wide range of involved energy and of size of the flares make possible to investigate the physical processes at the base of the storage and release of energy in the solar atmosphere at different scales. In particular, small flares have some advantages that allow studing some aspects which cannot be investigated in strong events. Usually flares characterized by a lesser emission of energy do not saturate the digital images and allow us observing in detail the topology of the involved magnetic systems, like the flare loop configuration or the ribbons shape. Moreover, this kind of flares has the advantage that can be observed entirely within the field of view of high resolution instruments. However, due to the unpredictable character of these events, it is not easy to acquire good quality datasets of flares in general and of small flares in particular.

One of the principal manifestations of the flare energy release is the presence of the ribbon brightenings in chromosphere, that are after observed in the ultraviolet (UV) wavelenghts. They are the effects of the accelerated particles from the reconnection site to the lower layers of the atmosphere. The location, the shape, and the motion of the ribbons provide useful information to understand the connectivity of the magnetic systems involved by the flare. Sometimes the ribbons show a peculiar shape which suggest some idea of the overlying topology of the magnetic field and an indication of the main reconnection site \citep[see, e.g.,][]{Sav15}. For example, circular ribbons are often interpreted as the evidence of the presence in corona of a three-dimensional (3D) null point \citep{Mas09, Rei12, Wan12, Sun13, Jia14, Man14, Zha15, Liu15}. Such a topological field model is associated to many flares, although it is not easy to observe this kind of ribbons, probably due to the asymmetry of the real configuration and to the privileged direction of particle acceleration. The presence of a 3D null point determines the formation of a surface, named fan, and two singular field lines, named inner and outer spine, belonging to two different connectivity domains \citep{Mas12}. Both the fan and the spines are sites where current sheets form (see below).

\citet{Mas09} observed a confined flare presenting a circular ribbon which brightens sequentially along the counterclockwise direction. In that case, the emergence of new magnetic field induced the injection of magnetic free energy, while the shearing of the spines relatively to the fan surface led
to a variation of the null-point geometry. From their simulation, they found that slipping reconnection \citep{Aul06} and traditional cut-and-paste reconnection occurred sequentially, being able to explain the observed features. \citet{Pon13}, simulating the effect of a spine-fan magnetic
reconnection process at a coronal null point by means of analytical and computational models, observed that the flipping of magnetic field lines in a manner similar to that observed in quasi-separatrix layers or the slip-running reconnection may occur in such a magnetic configuration. 

The slipping/slip-running reconnection in a fan-spine magnetic topology seems to be also associated with the occurrence homologous jets, as observed by \citet{Wan12}, and the multiple-ribbon formation \citep[e.g.,][]{Gug16}. 

Moreover, the formation of a null point topology has been invoked in more complex phenomena, like the formation and eruption of an active region (AR) sigmoid \citep{Jia14} or the violent events occurring in a cluster of ARs \citep{Man14}. Recenty, \citet{Liu15} observed a circular ribbon as a consequence of a filament eruption. They analyzed a GOES-class X1.0 flare occurred on 2014 March 29 in AR NOAA 12017 where the asymmetric eruption of a sigmoidal filament seemed to be initiated due to a magnetohydrodynamics (MHD) instability which disturbed the fan-spine-like field causing the breakout-type reconnection at a coronal quasi-separatrix layer.

The magnetic reconnection in a 3D null point has been also studied in several numerical simulations \citep{Gal97, Gal03, Pon07a, Pon07b, Par09}. \citet{Pon07b} showed that rotations of the spine produce current sheets more extended along the fan, and viceversa. The shearing motions at the fan and spine footpoints are also source of modification of the 3D null point configuration and of the subsequent formation of current sheets \citep{Pon07a}. In this context, \citet{Par09} observed that reconnection cannot occur in an axisymmetric null-point topology when the boundary driving is also axisymmetric around the spine footpoint. 

Computing a potential field extrapolation and applying horizontal motions observed by SOHO/MDI to the photospheric boundary of the computational box, \citet{Bau13a} and \citet{Bau13b} performed some MHD simulations to investigate the particle acceleration in a 3D null point region and their deposition. From studies of the locations of the non-thermal electrons and of their acceleration paths, they confirmed that the energy density of accelerated particles close to the null point has a peak, while ribbons associated with the outer spine spread much more, decreasing their energy density and producing elongated flare ribbons \citep{Pon16}.

More recently, \citet{Wyp16}, taking into account that coronal jets can be described by a 3D null point configuration, found that the longest-lasting and most energetic events occur when there is a large ratio N/L, where N is the size of the jet source region, corresponding to the fan size, and L is the size of the coronal loop that envelops the jet source, corresponding to the spine size. 

However, the conditions which lead to the occurrence of solar flares in such configuration have been rarely studied so far due to the difficulties to observe these topologies with enough spatial resolution.

In this Paper we analyze recent high resolution observations of a GOES B-class flare characterized by a symmetric magnetic field configuration  producing a closed circular ribbon at the chromospheric level. This event provides a further contribution to the few observations of such a kind of ribbon reported in literature and confirms the results of some models proposed for the description of the jetting activity observed in corona. In the next Section we describe the whole dataset and its analysis, while in Section 3 we show the results. Finally, in Section 4 we summarize the main conclusions.

\section{Observations and Data analysis}

The target of this study is AR NOAA 12351, which was observed on 2015 May 20, from 14:21 UT to 14:31 UT and from 15:09 UT to 15:29 UT using high temporal, spatial and spectral resolution data acquired by the Interferometric BIdimensional Spectrometer \citep[IBIS;][]{Cav06} instrument and the ROSA imager \citep{Jes10} operating at the NSO/Dunn Solar Telescope (DST). During the second time interval the AR was located at about N22 E42 and a GOES B-class flare occurred from 15:03 UT to 15:30 UT.

The IBIS data set consists of 30 scans of two photospheric lines in full polarimetric mode (\ion{Fe}{1} 630.25 nm and \ion{Fe}{1} 617.3 nm lines) and two chromospheric lines without polarimetric measurements (\ion{Ca}{2} 854.2 nm and H$\alpha$ 656.28 nm lines). All of these lines were sampled with a FWHM of 2 pm, an average step of 2 pm, and an integration time of 60 ms. The sampling was of 30 spectral points along the \ion{Fe}{1} 630.25 nm line, 24 along the \ion{Fe}{1} 617.3 nm line, 25 along the \ion{Ca}{2} 854.2 line, and 17 along the H$\alpha$ 656.28 nm line. The time spent for each scan was about 67 sec. The field of view (FOV) was 500 $\times$ 1000 pixels with a pixel scale of 0\farcs095. For each spectral frame, a simultaneous broad-band (at $633.32 \pm 5$ nm) frame, imaging the same FOV with the same exposure time, was acquired. To reduce the seeing degradation, the images were restored using the Multi-Frame Blind Deconvolution \citep[MFBD;][]{Lof02} technique (see details in \citealp{Rom13}). At the same time of the IBIS data acquisitions, ROSA took images at the H$\beta$ 486.1 nm channel, with a FOV of 512 $\times$ 512 pixels, a pixel scale of 0\farcs095, and a time cadence of 0.25 s.

To determine the evolution of the intrinsic magnetic field strength we performed a single-component inversion of the Stokes profiles for all of the scans of the \ion{Fe}{1} lines using the SIR code \citep{Rui92}. We inverted the Stokes profiles of both \ion{Fe}{1} lines separately. For further details about the inversion method we refer to \citet{Rom13}. 

To compute the line-of-sight (LOS) velocity fields we reconstructed the profiles of the \ion{Ca}{2} 854.2 line in each spatial pixel by fitting the corresponding Stokes I component with a linear background and a Gaussian shaped line. The values of LOS velocity have been deduced from the Doppler shift of the centroid of the line profiles in each spatial point with respect to the median of the centroid in the whole FOV. We estimated the uncertainty affecting the velocity measurements by considering the standard deviation of the centroids of the line profiles estimated in all points of the whole FOV. Thus, the estimated error in the velocity is $\pm$0.2 km s$^{-1}$.

We also used the Space-weather Active Region Patches (SHARPs) data \citep{Hoe14} acquired by HMI/SDO \citep{Sch12} in the \ion{Fe}{1} 617.3 nm line on 2015 May 20, from 00:00 UT to 23:48 UT with a pixel size of 0\farcs51 and a time cadence of 12 min.

\section{Results}

\begin{figure}
\begin{center}
\includegraphics[trim=10 155 80 430, clip, scale=0.8]{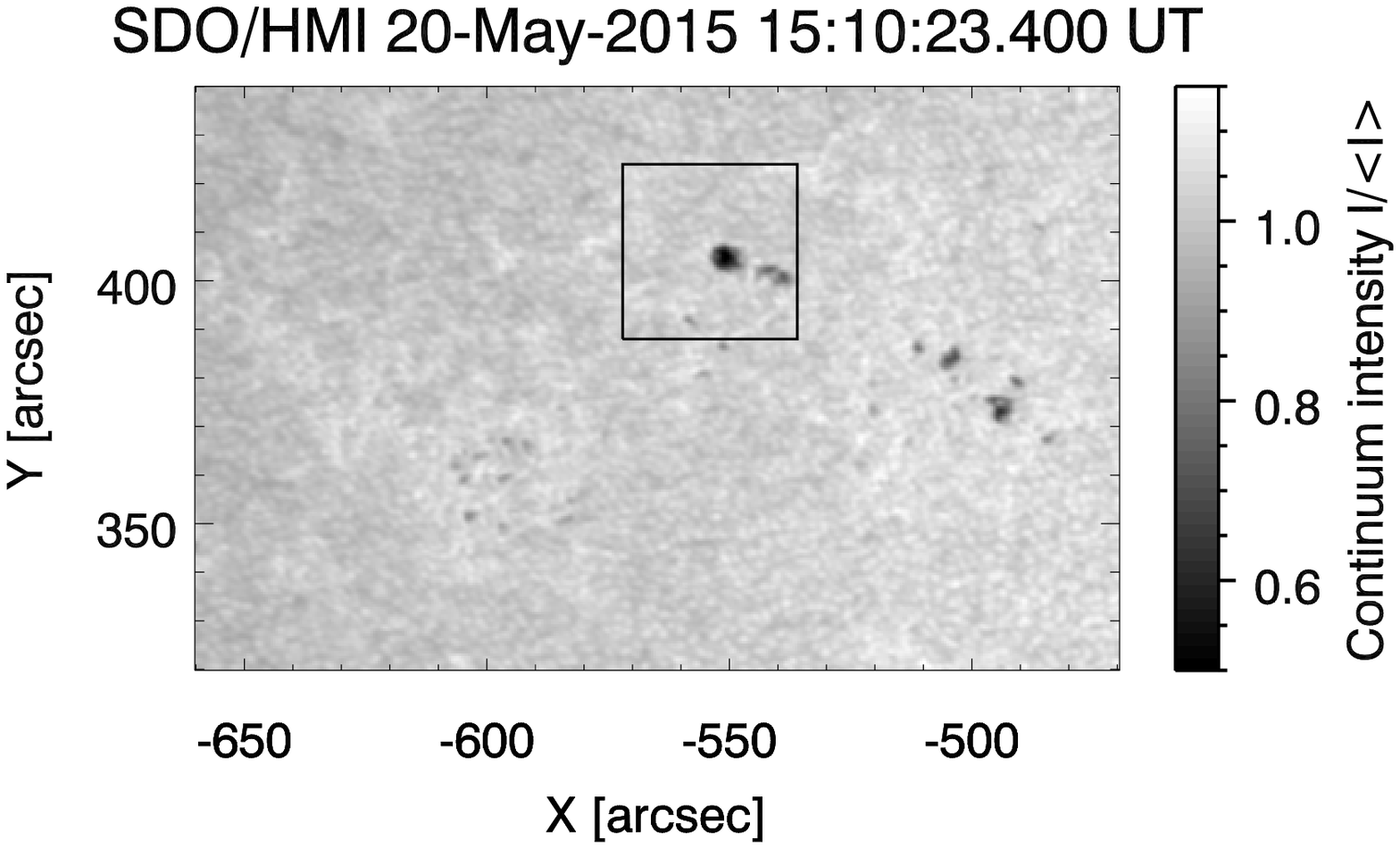}\\
\includegraphics[trim=10 100 80 430, clip, scale=0.8]{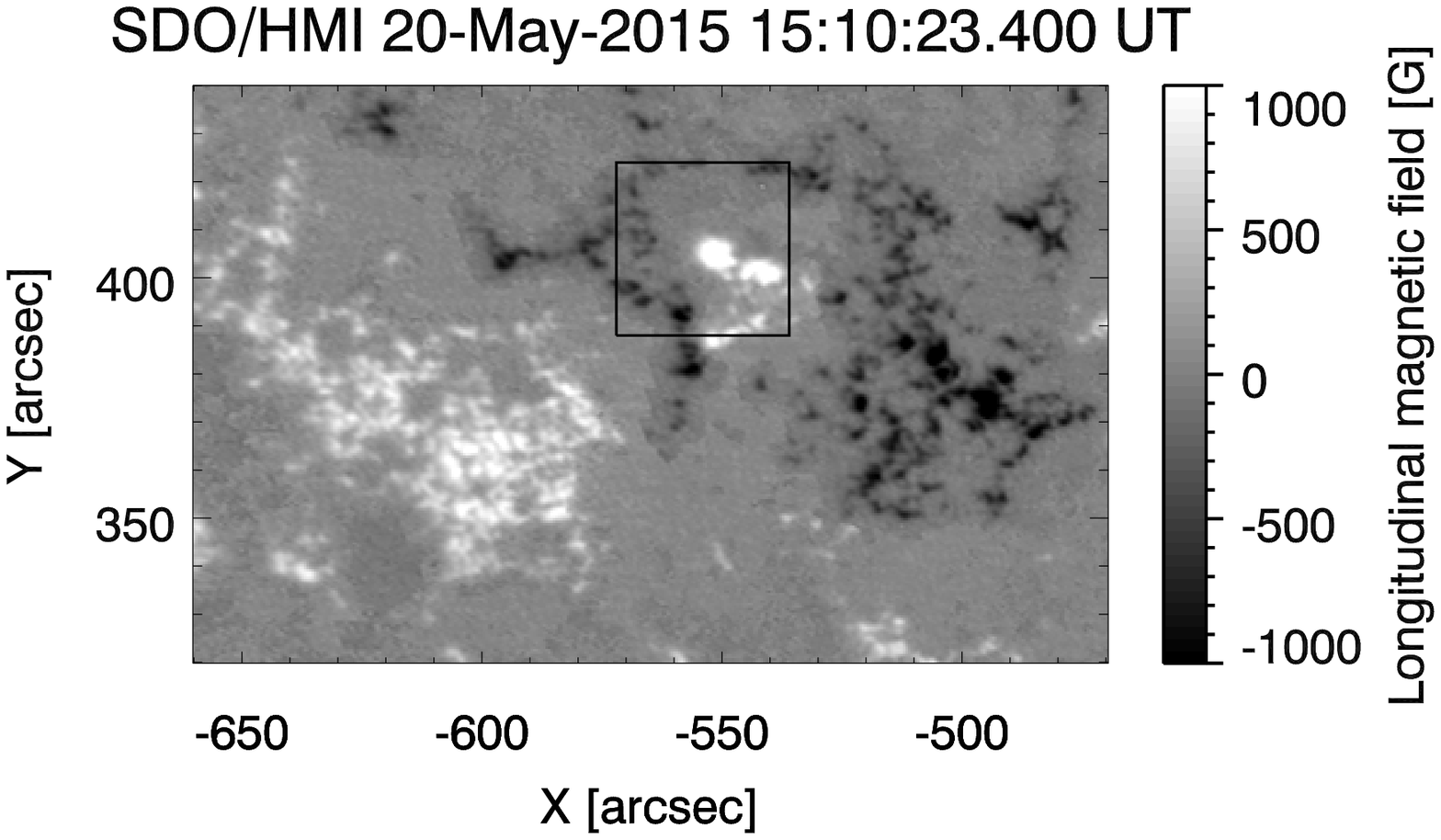}
\caption{SHARPs continuum image ({\em top panel}) and magnetogram ({\em bottom panel}) of the AR NOAA 12351 acquired by HMI/SDO. The black boxes indicate the IBIS FOV.} 
\label{Fig1}
\end{center}
\end{figure}

On 2015 May 20, AR NOAA 12351 was characterized in the photosphere by some pores ({\em top panel} of Figure \ref{Fig1}) and a quite diffuse magnetic field ({\em bottom panel} of Figure \ref{Fig1}), with the preceding and the following polarities negative and positive, respectively.

The IBIS FOV was located over a pore of the AR characterized by positive magnetic field, as indicated by the black boxes in Fig. \ref{Fig1}. 

\begin{figure}
\begin{center}
\includegraphics[trim=10 200 10 200, clip, scale=0.5]{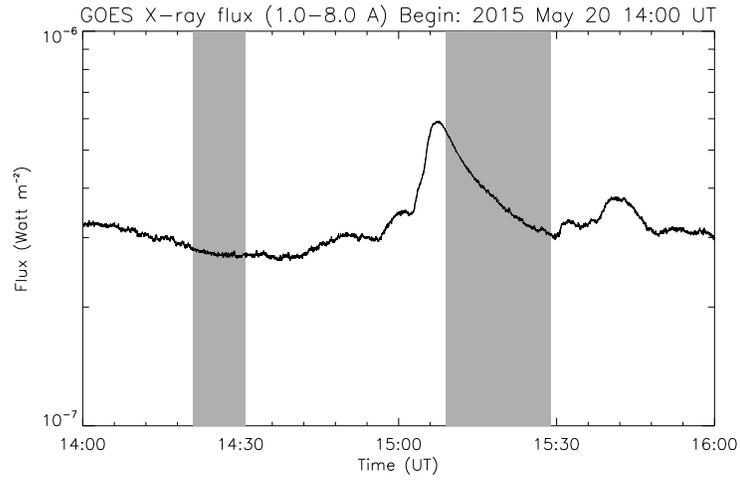}
\caption{X-ray profile between 1.0 and 8.0 \AA{} registered by GOES. The grey bands indicate the time windows when DST was observing.} 
\label{Fig2}
\end{center}
\end{figure}

As we can see from Figure \ref{Fig2}, where the X-ray GOES profile integrated between 1.0 and 8.0 \AA{} and the observational time windows of the DST are displayed, high resolution photospheric and chromospheric data were available before the flare and during its main phase. 

\begin{figure}
\begin{center}
\includegraphics[trim=0 165 200 430, clip, scale=0.55]{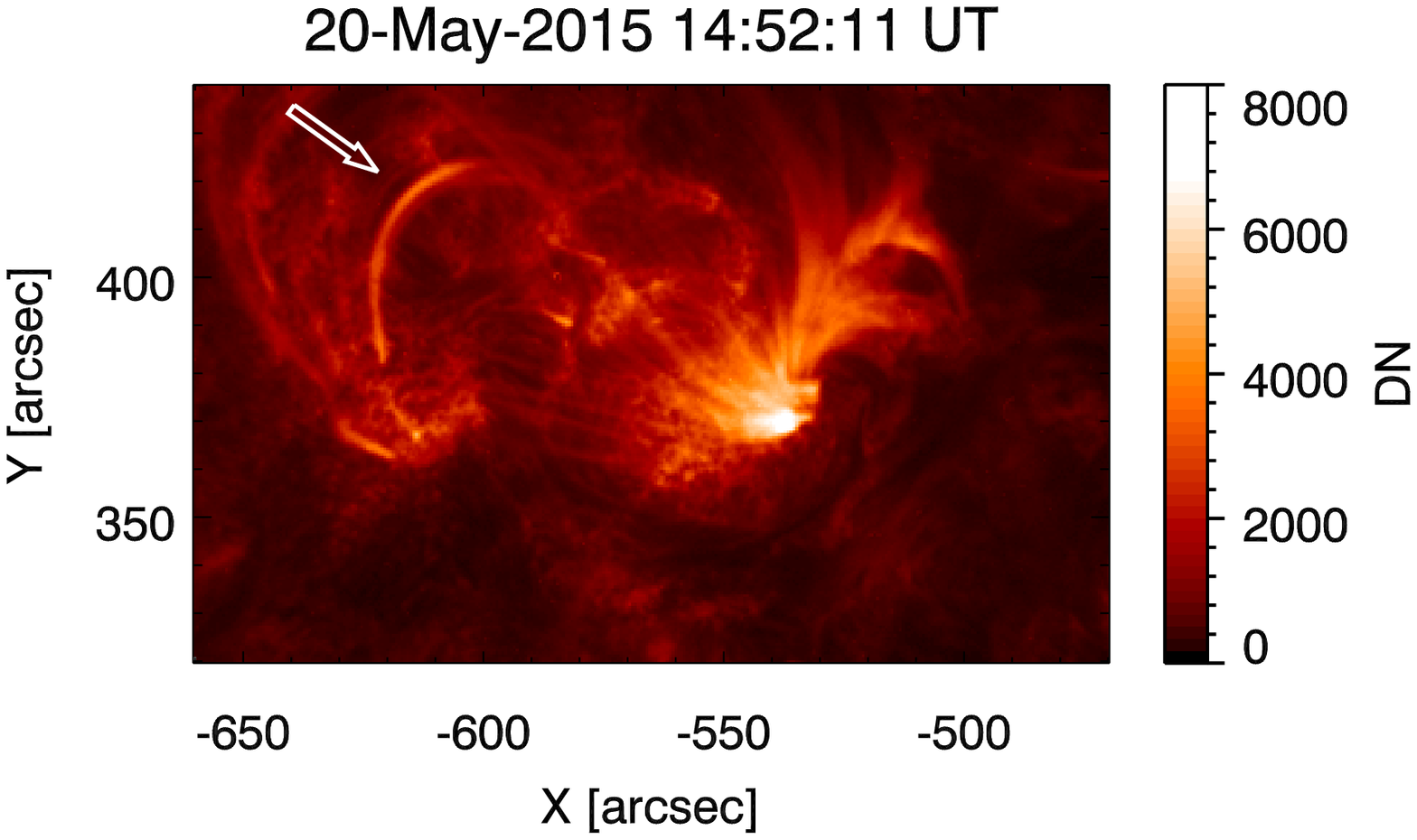}
\includegraphics[trim=70 165 90 430, clip, scale=0.55]{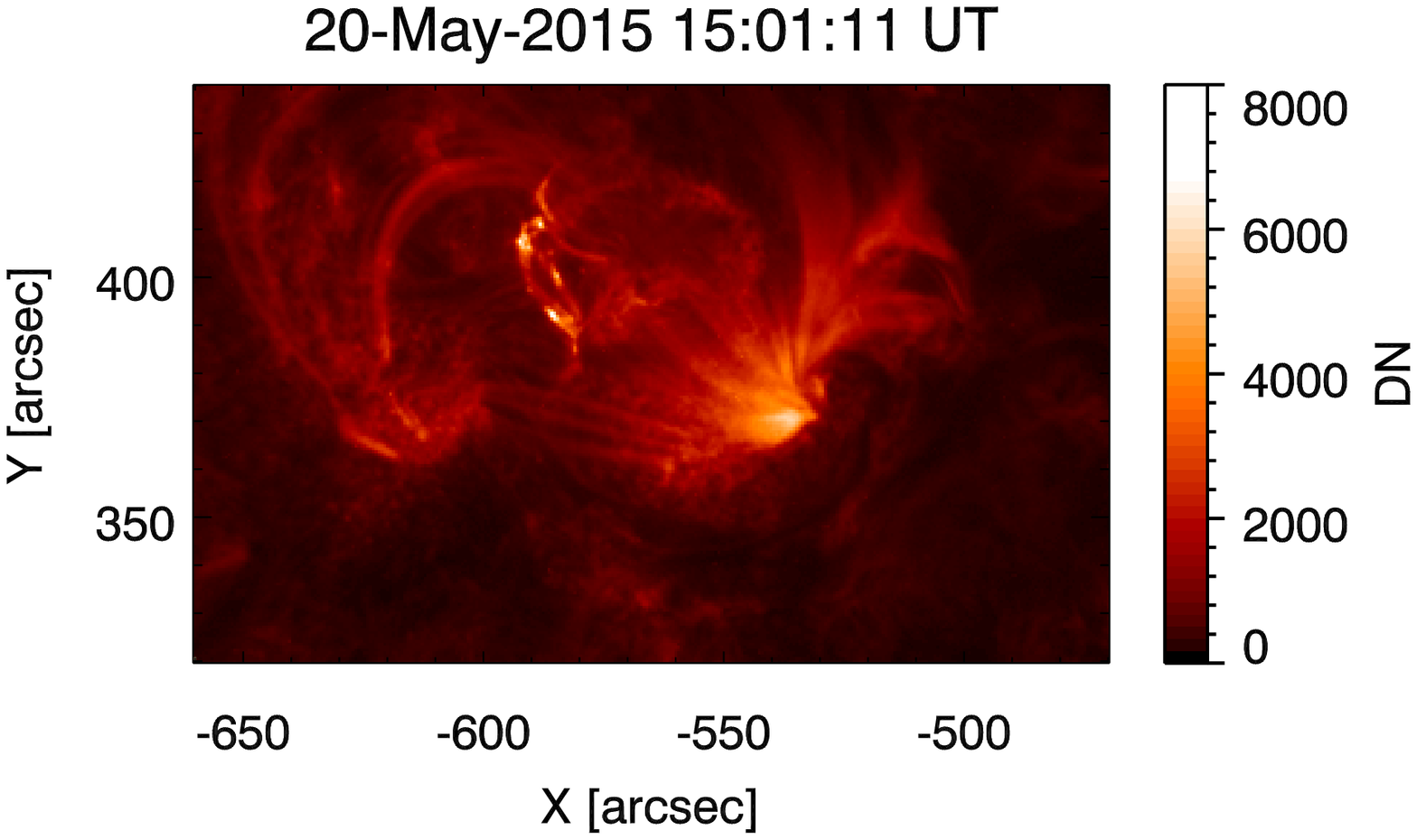}\\
\includegraphics[trim=0 105 200 430, clip, scale=0.55]{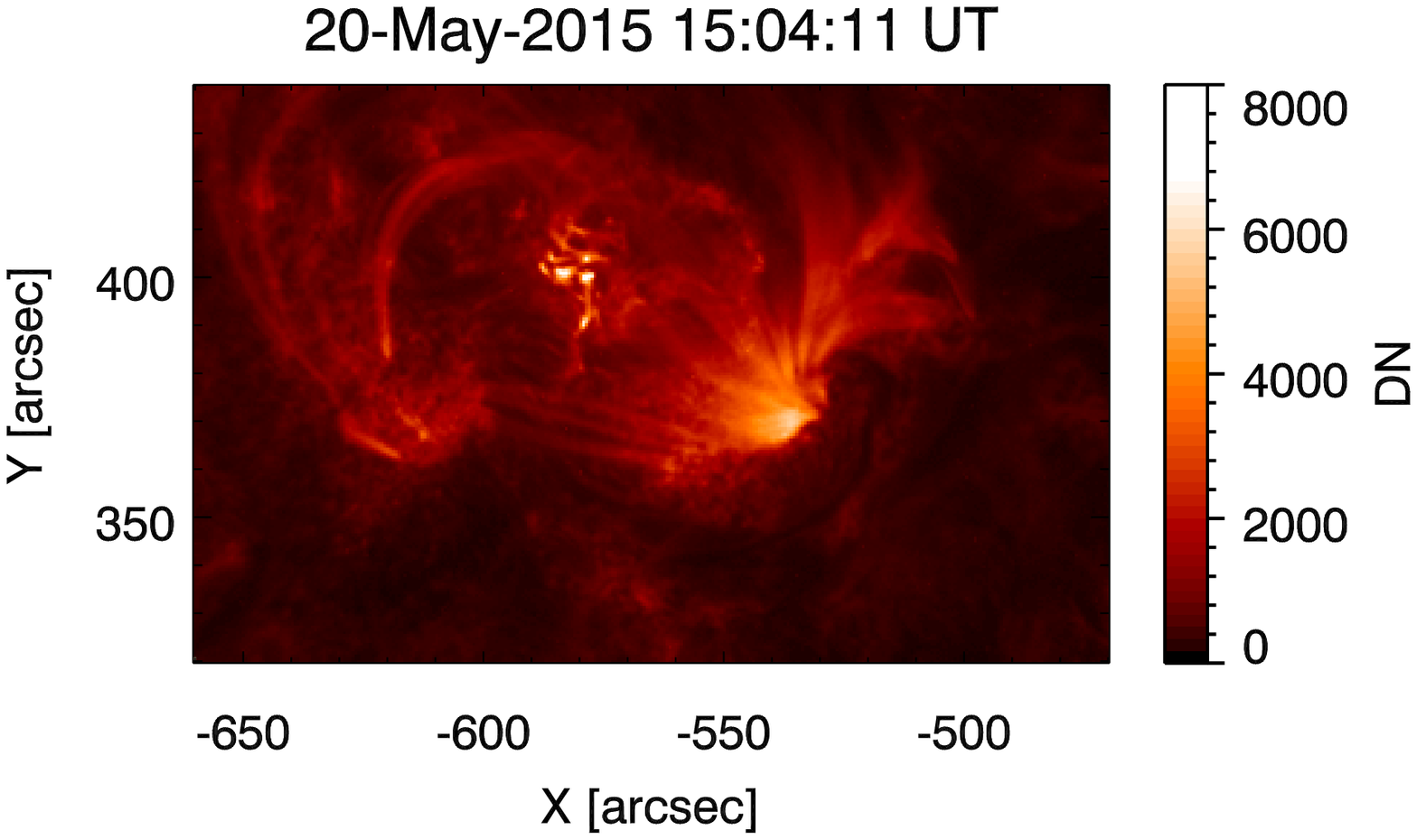}
\includegraphics[trim=70 105 90 430, clip, scale=0.55]{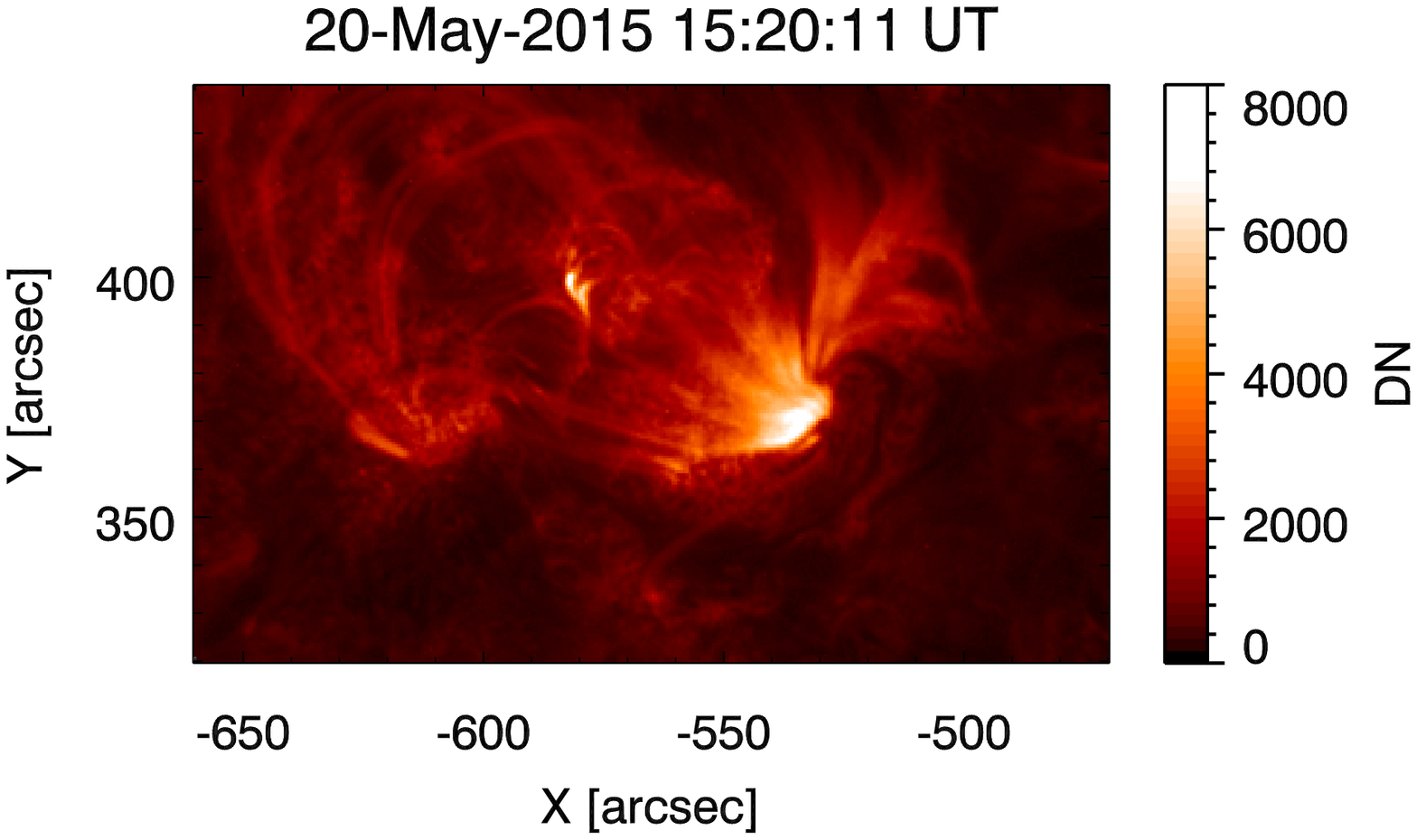}\\
\caption{Sequence of flare observed in corona by AIA/SDO at 171 \AA{}. (An animation of this figure is available in the online journal.)} 
\label{Fig2bis}
\end{center}
\end{figure} 

\begin{figure}
\begin{center}
\includegraphics[trim=20 250 30 250, clip, scale=0.6]{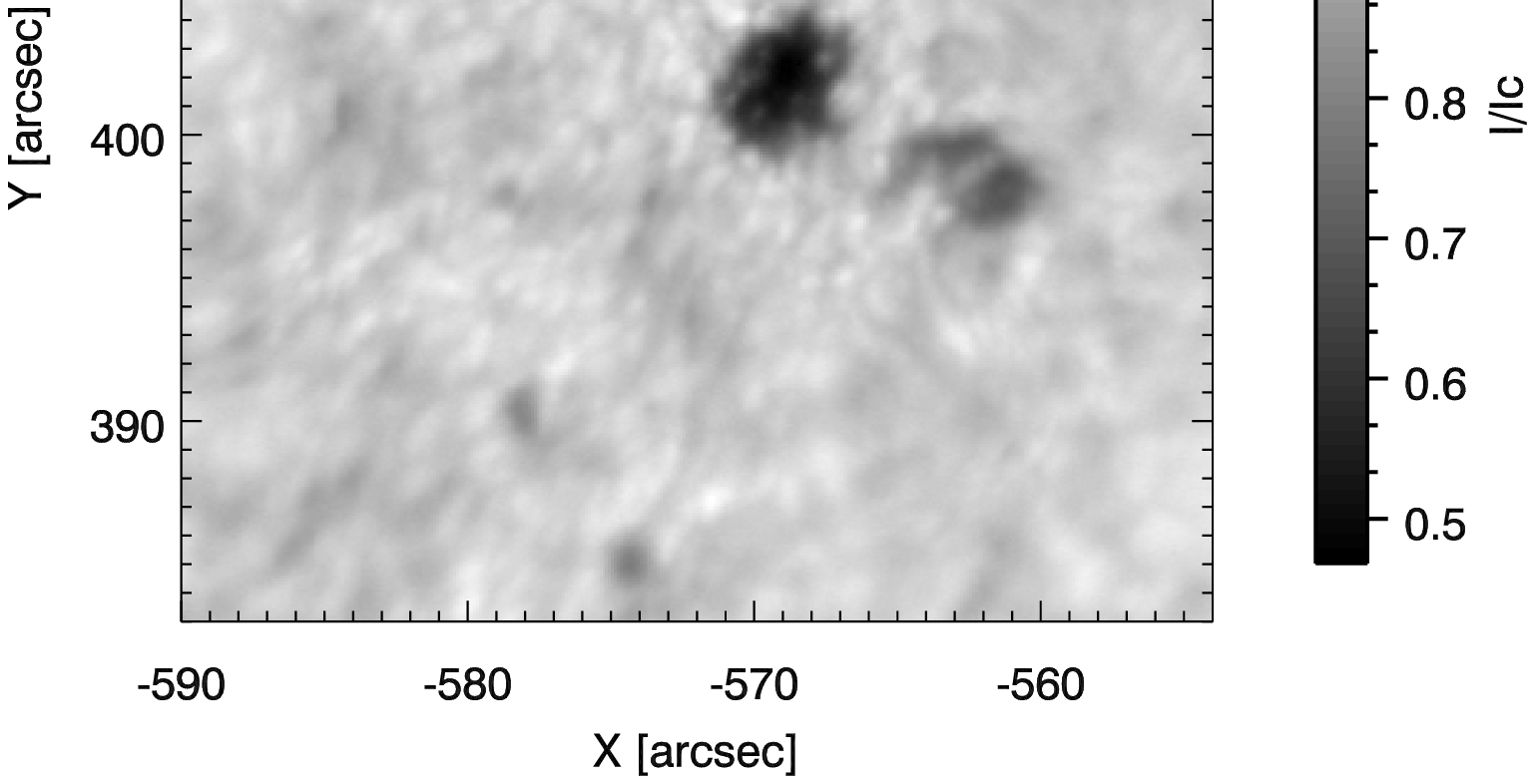}
\includegraphics[trim=20 210 30 250, clip, scale=0.6]{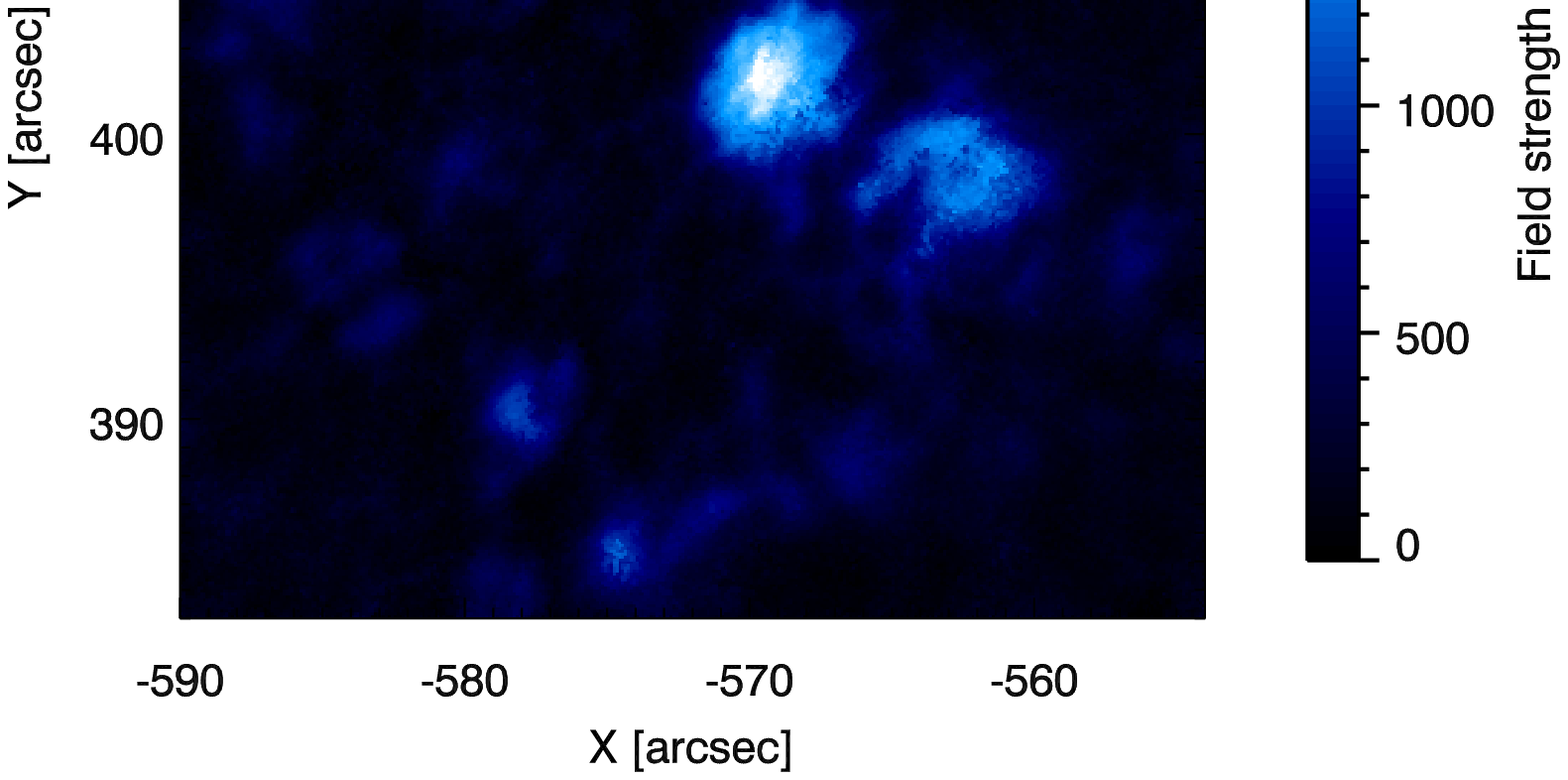}
\caption{High resolution observations of the main pore of AR NOAA 12351 acquired by IBIS. Continuum ({\em top panel}) and magnetic field intensity ({\em bottom panel}) taken during the main fase of the flare.} 
\label{Fig3}
\end{center}
\end{figure}

The flare evolution is well described by the images of the corona taken by AIA/SDO at 171 \AA{} (Figure \ref{Fig2bis} and the corresponding online movie). During the pre-flare phase we note a bright loop connecting the western part of the AR with its eastern part (see the arrow in the {\em top left} panel of Figure \ref{Fig2bis}). The flash-phase of the flare corresponds to the brightening of a small but complex bundle of loops located at [-580\arcsec$,$400\arcsec] (see the {\em top right} and {\em bottom left} panels of Figure \ref{Fig2bis}). While the main phase of the flare does not show any significant emission at 171 \AA{}, as shown in the {\em bottom right} panel of Figure \ref{Fig2bis}.

The high resolution image of the main pore of AR NOAA 12351, taken in the center of the \ion{Fe}{1} 630.25 nm line during the main phase of the flare ({\em top panel} of Figure \ref{Fig3}), shows that the seeing conditions were not very good at that time, even if it is possible to see other few smaller pores around. The magnetic field intensity in the main pore reaches 2000 G, while others concentration of the magnetic field above 1000 G correspond to the other smaller pores ({\em bottom panel} of Figure \ref{Fig3}). 

\begin{figure}
\begin{center}
\includegraphics[trim=10 130 200 350, clip]{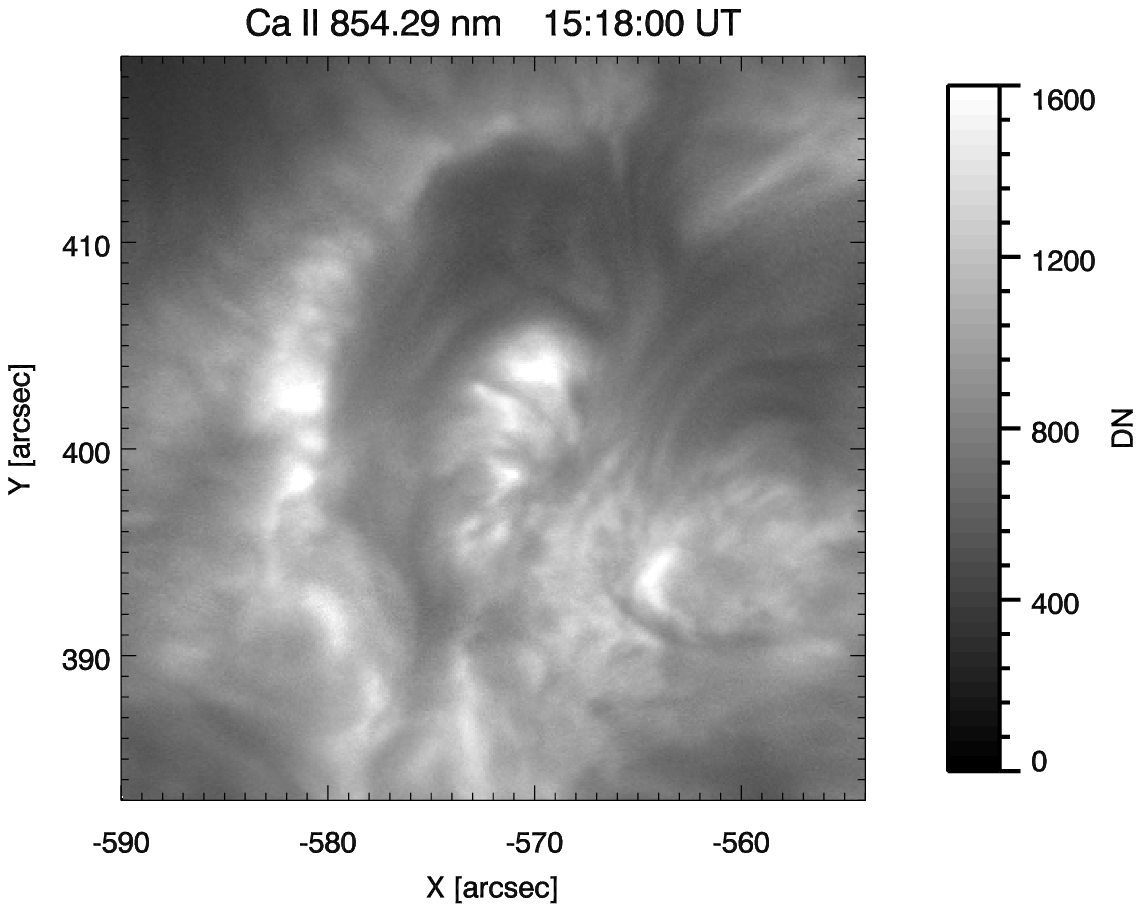}
\caption{Observation of the same FOV of Figure \ref{Fig3} in the center of the \ion{Ca}{2} 854.2 line.} 
\label{Fig4}
\end{center}
\end{figure}

\begin{figure}
\begin{center}
\includegraphics[trim=10 152 100 320, clip, scale=0.42]{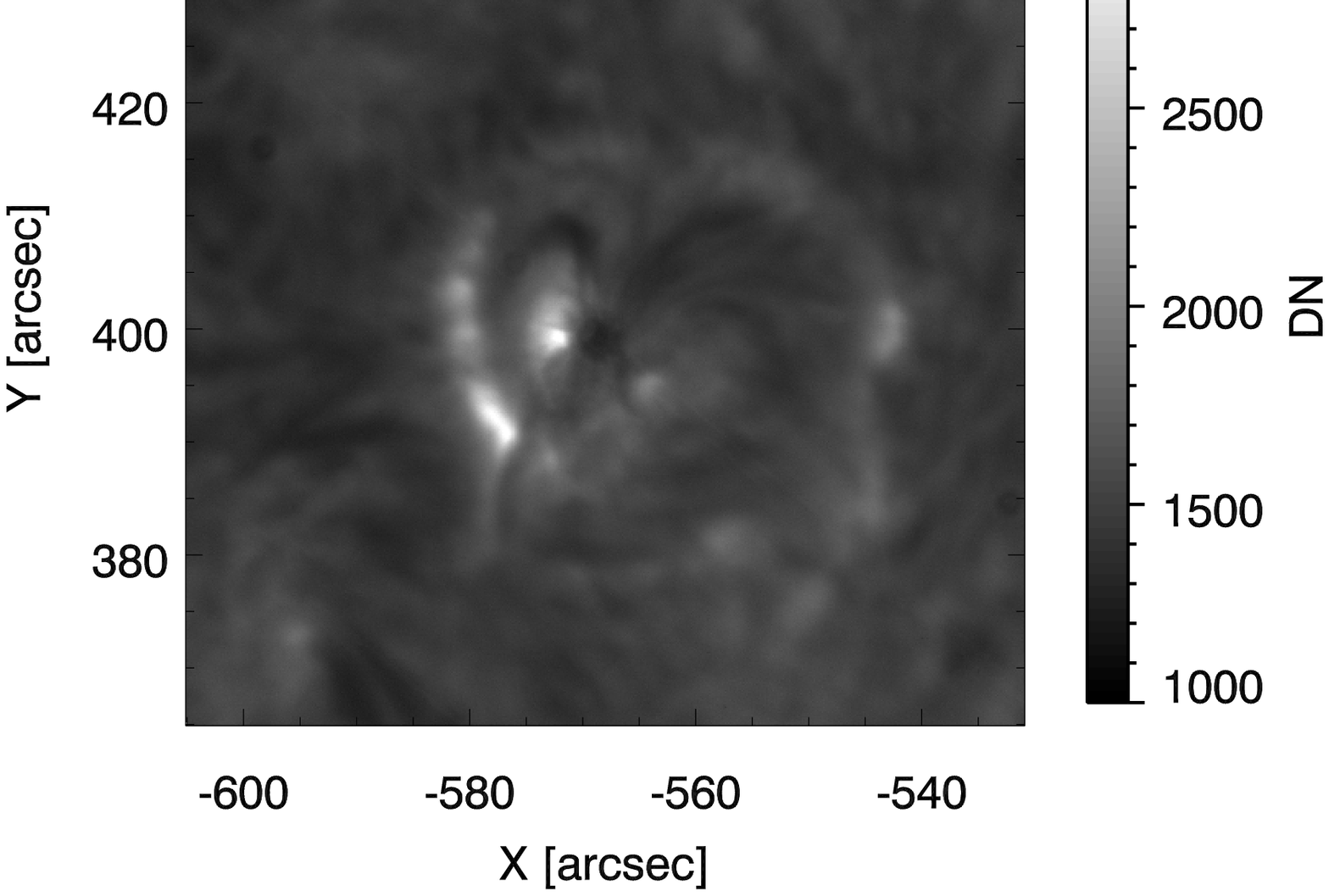}
\includegraphics[trim=72 152 100 320, clip, scale=0.42]{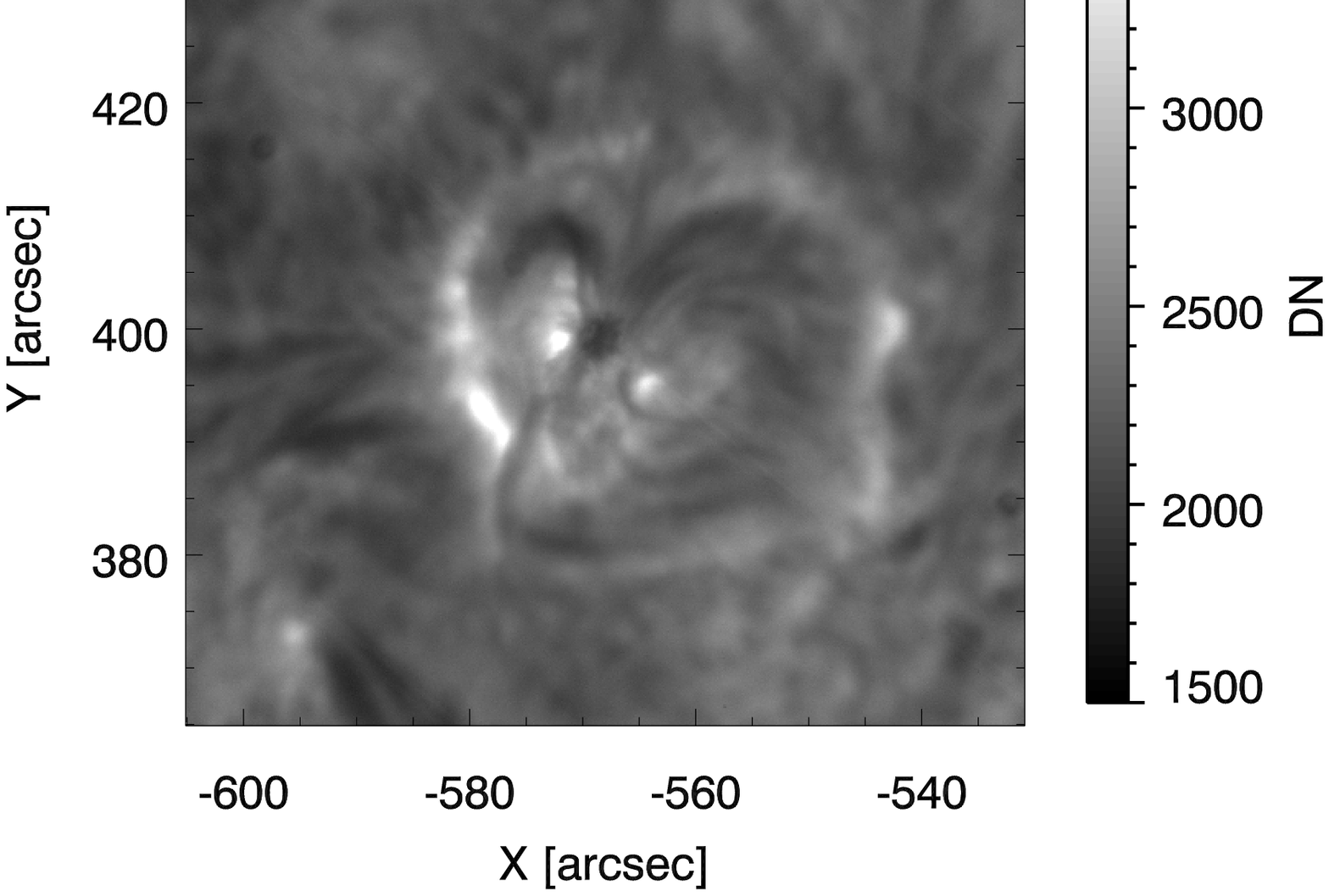}\\
\includegraphics[trim=10 152 100 320, clip, scale=0.42]{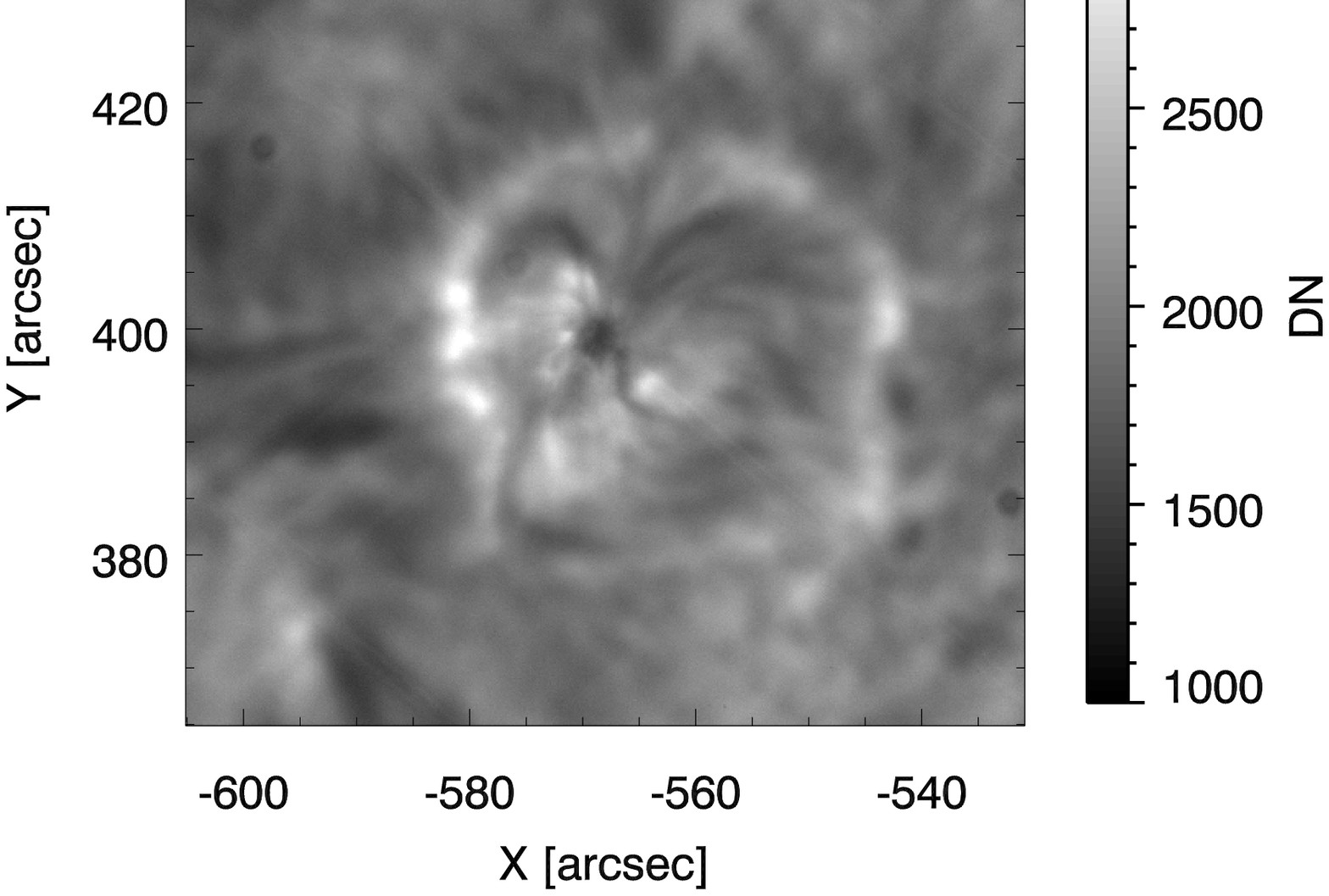}
\includegraphics[trim=72 152 100 320, clip, scale=0.42]{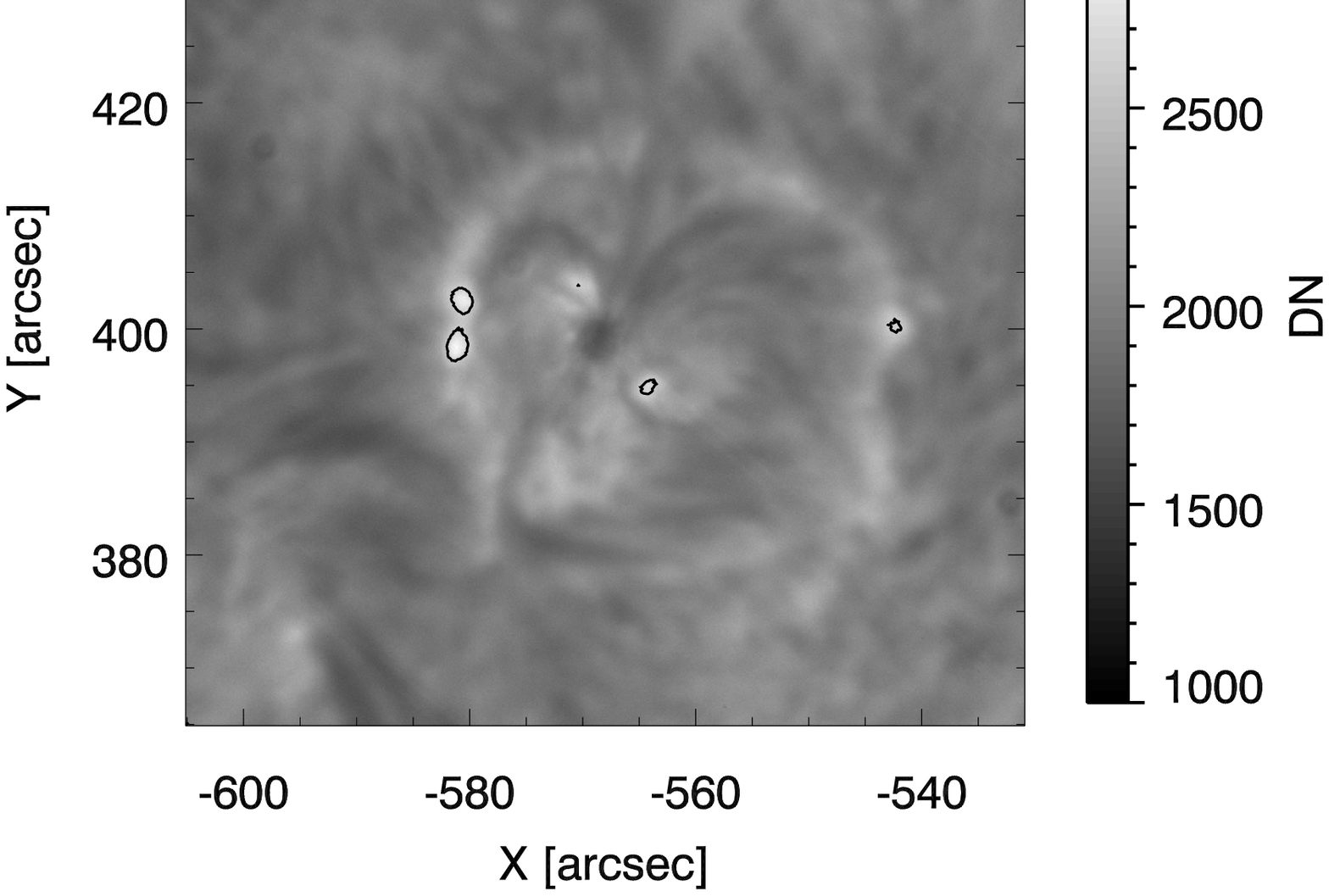}\\
\includegraphics[trim=10 110 100 320, clip, scale=0.42]{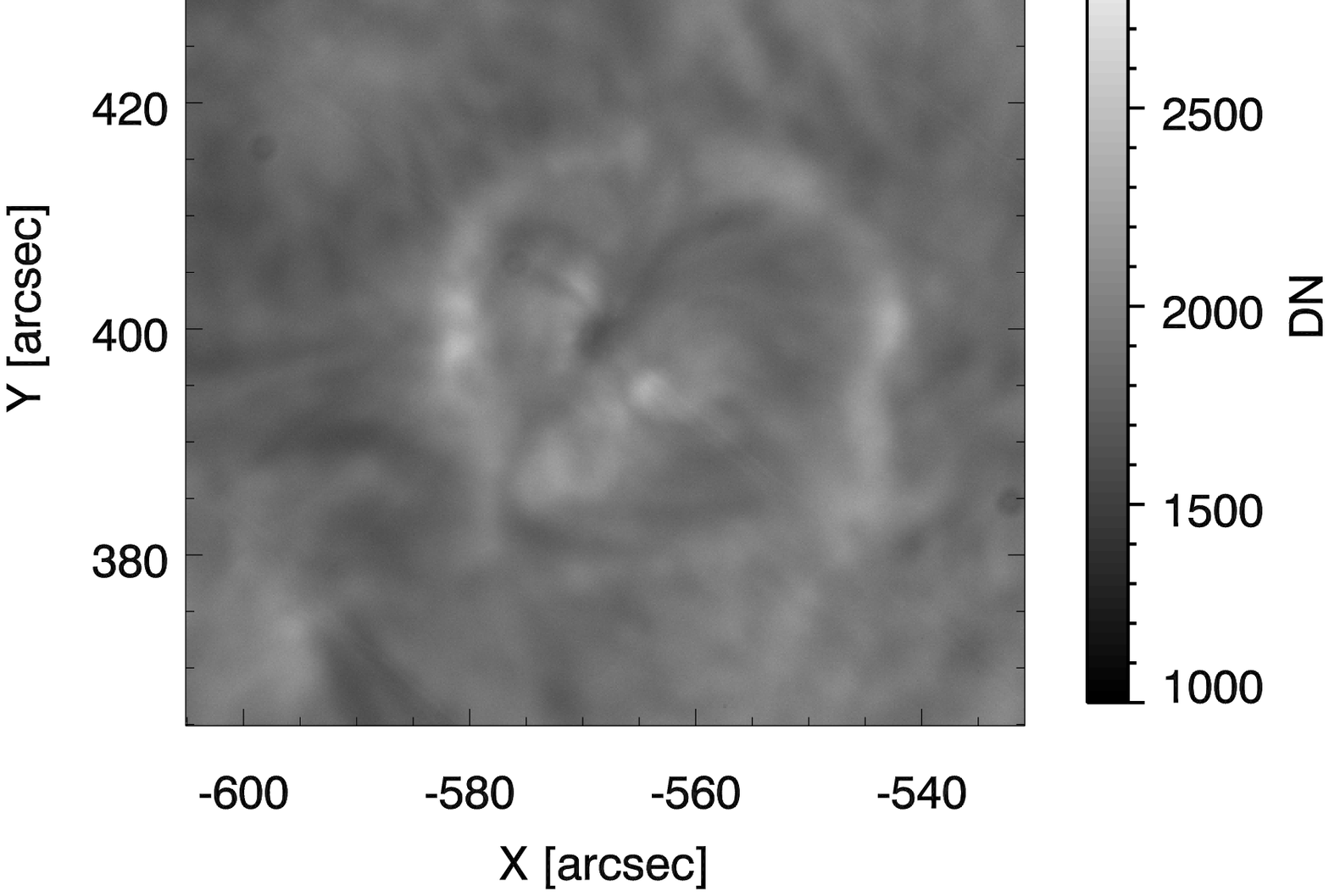}
\includegraphics[trim=72 110 100 320, clip, scale=0.42]{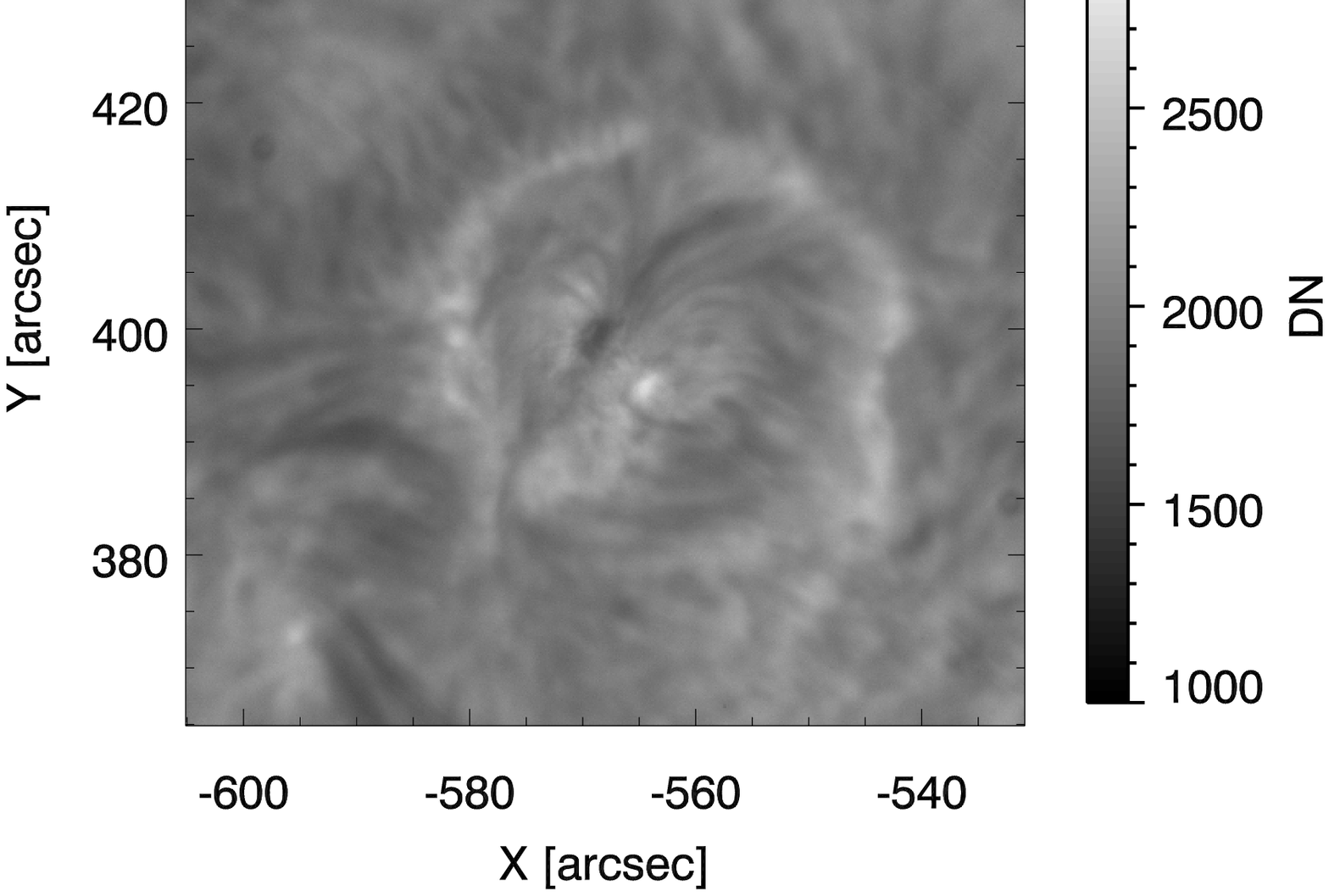}\\
\caption{Evolution of the semi-circular ribbon observed by ROSA at 486.1 nm. The black contours in the {\em right middle panel} indicate the position of the brighter regions at 15:19:41 UT.} 
\label{Fig5}
\end{center}
\end{figure}

During the main phase of the flare we observe in the IBIS FOV parts of the ribbons at chromospheric level. At 854.2 nm we distinguish some small filaments connecting the main pore, visible in the center of the FOV, to the eastern surrounding ribbons (see Figure \ref{Fig4}). However, the IBIS FOV does not allow to see the whole surrounding ribbon and its shape. For this reason we use the ROSA imager which allows us to observe the ribbon evolution in H$\beta$. At this wavelength we see more clearly that the filaments are distributed radially from the main pore. Their outer footpoints describe a circle around the pore. During the main phase of the flare, these footpoints start to brighten sequentially in clockwise direction from the eastern to the western ones, producing a semi-circular ribbon with a diameter of about 40\arcsec{} and formed by several adjacent bright kernels of 1\arcsec-2\arcsec, as shown in Figure \ref{Fig5}. In particular, at 15:09 UT during the main phase of the flare, when we started our observations, we see that the brightenings are more intense at the outer footpoints of the darker filaments ({\em top left panel} of Fig. \ref{Fig5}). Although we were not able to observe the peak of the flare, we see also some brightenings near the main pore, i.e., corresponding to the inner footpoints of these filaments.

\begin{figure}
\begin{center}
\includegraphics[trim=10 100 0 200, clip, scale=0.7]{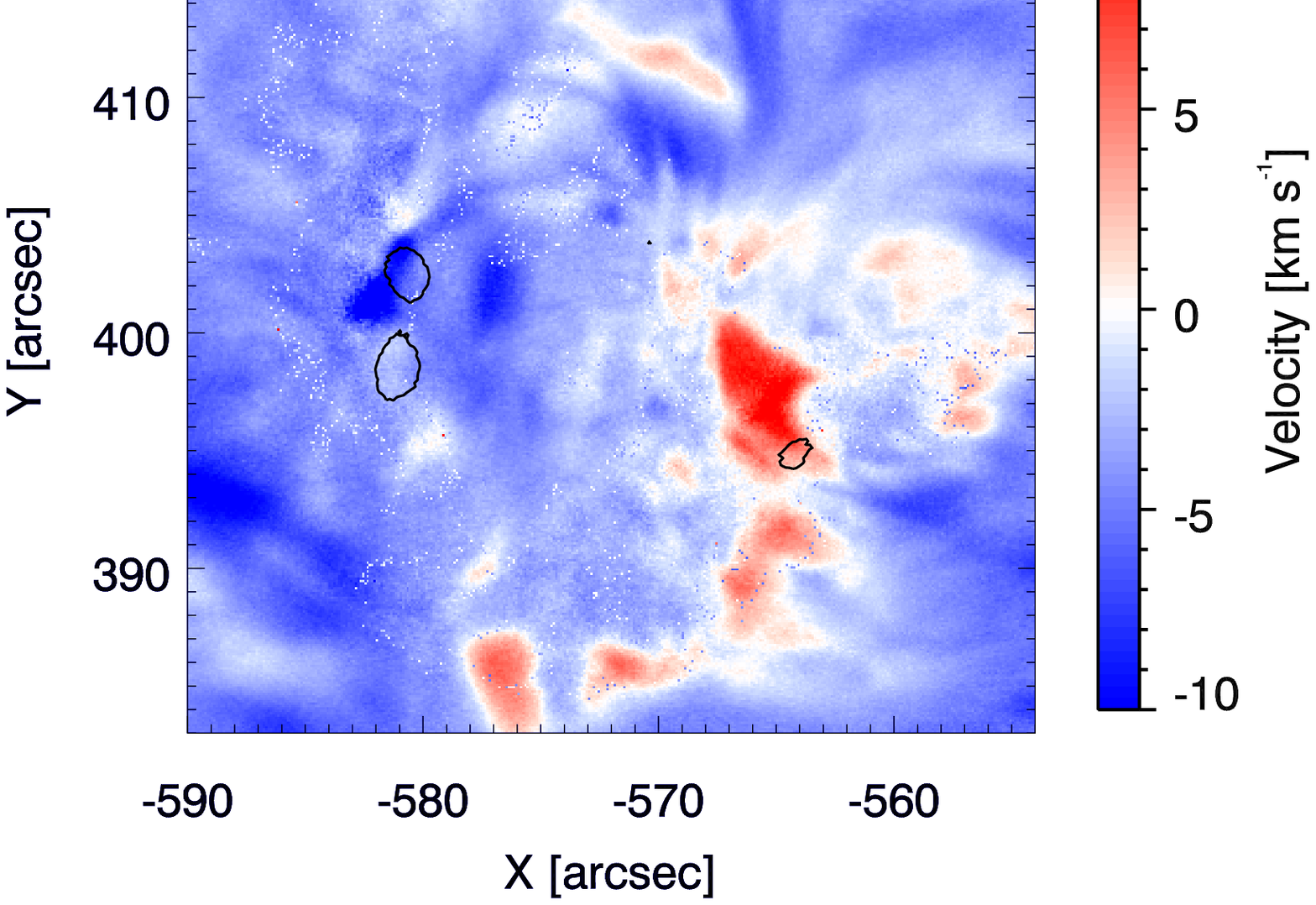}
\caption{LOS velocity map obtained by the Doppler shift of the centroid of the \ion{Ca}{2} line. Positive (negative) values indicate motions away from (towards) the observer. The black contours indicate the position of the brighter regions in the H$\beta$ line at 15:19:41 UT (see the {\em right middle panel} of Figure \ref{Fig5}).} 
\label{Fig6}
\end{center}
\end{figure}

The Doppler shift of the centroid of the \ion{Ca}{2} line shows some interesting features at the chromospheric level. In the LOS velocity field map (Figure \ref{Fig6}) we recognize that the darker filaments are characterized by a motion of the plasma towards the observer more intense than the background and a blob of plasma, located at [-567\arcsec$,$397\arcsec], which is moving away from the observer. This blob has the same location of the brightening near the main pore visible at the same time in the center of H$\beta$ line (compare to the {\em right middle panel} of Figure \ref{Fig5}). The other smaller regions characterized by plasma moving away from the observer correspond to the smaller pores visible in the bottom edge of the IBIS FOV (see the {\em left panel} of Figure \ref{Fig3}). 

\begin{figure}
\begin{center}
\includegraphics[trim=0 0 0 0, clip, scale=0.3]{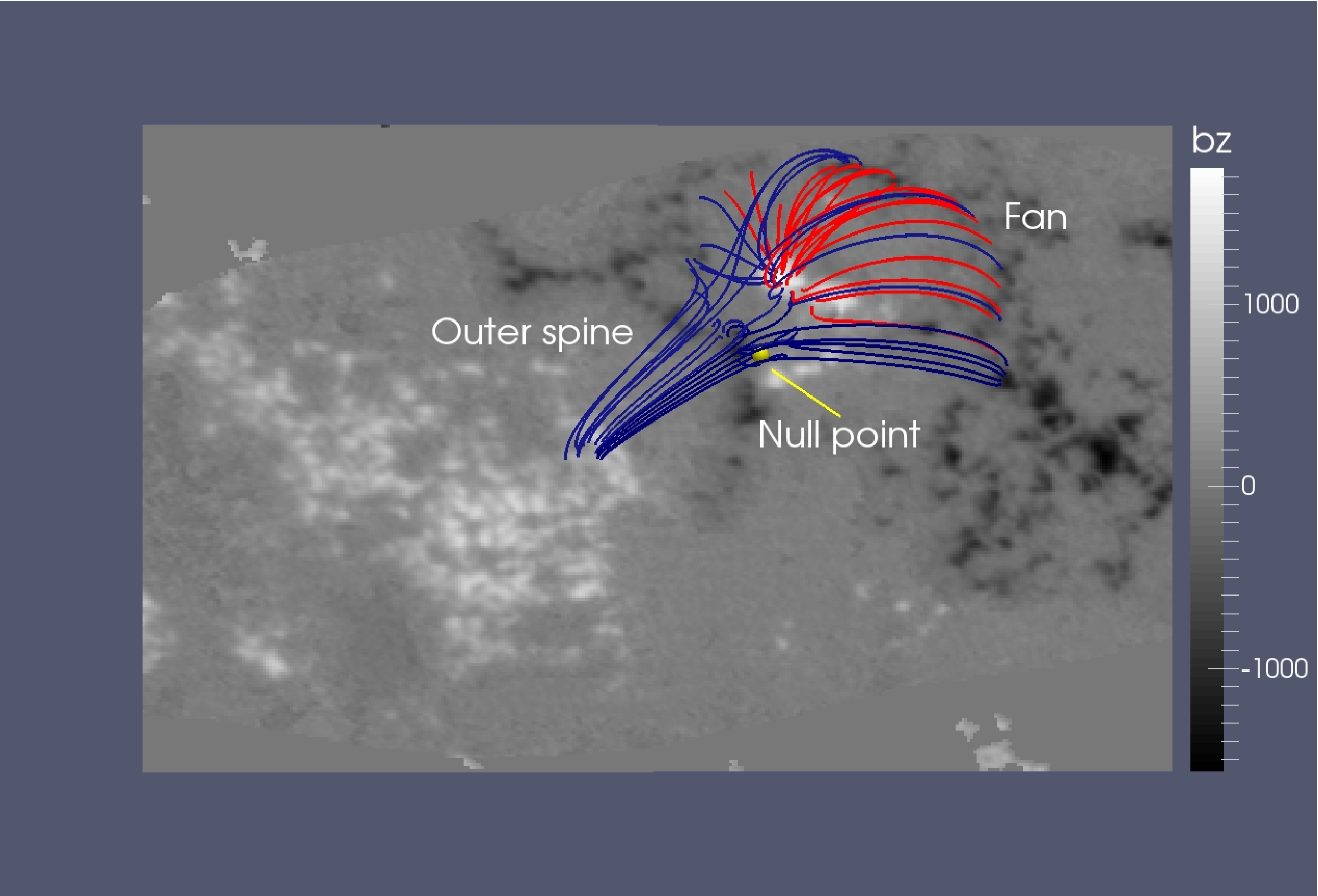}\\
\includegraphics[trim=0 0 0 0, clip, scale=0.3]{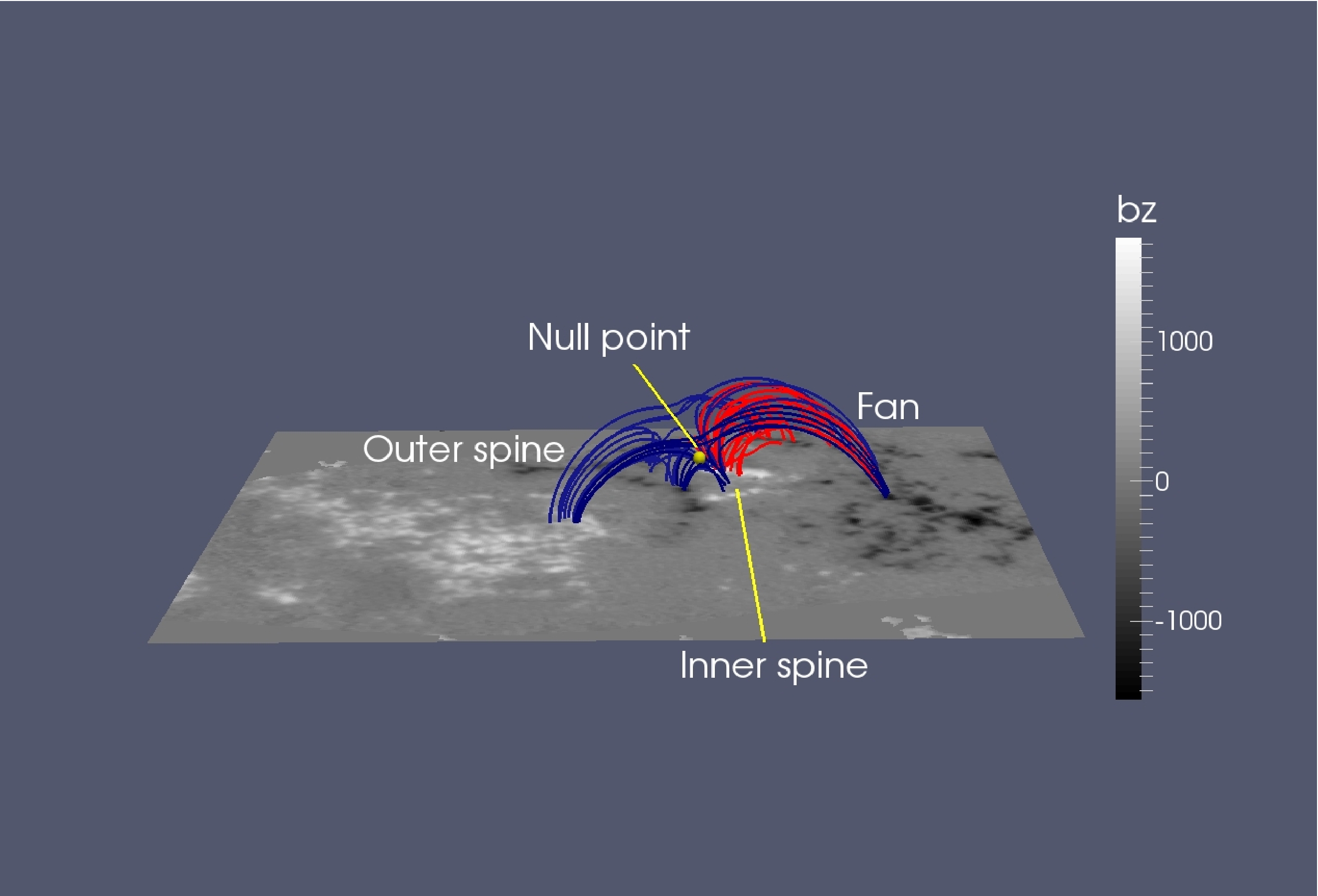}
\caption{Top ({\em top panel}) and side ({\em bottom panel}) views of the potential field extrapolation obtained by means of the magnetogram acquired by HMI/SDO on May 20 at 14:22 UT used as photospheric boundary conditions.} 
\label{Fig7}
\end{center}
\end{figure}

In order to get an idea of the magnetic configuration underlying this small flare, leading for an explanation of the above mentioned observational results, we performed a potential field extrapolation using the method described by Alissandrakis (1981). We considered the magnetogram acquired by HMI/SDO on May 20 at 14:22 UT as photospheric boundary conditions. We also applied the method outlined in \cite{Bev06} to find the null points above the photospheric surface. The results are shown in Figure \ref{Fig7}.

We found a null point located above the region of interface between the positive patches, corresponding to the smaller pores in the southern part of IBIS FOV, and the negative network of the supergranular cell (see the {\em top panel} of Figure \ref{Fig7}). The presence of this null point determines the formation of a fan (red lines in Figure \ref{Fig7}) whose inner part seems to connect the main positive patch, corresponding to the main pore, with the northern part of the surrounding network, where the northern portion of the circular ribbon was located. From the side view of the extrapolation ({\em bottom panel} of Figure \ref{Fig7}) we also see that the footpoint of the inner spine is rooted around the main pore, where we observed some brightenings during the main phase of the flare. Instead, the outer spine is rooted in the positive magnetic field located to the south-western part of the AR, thus outside the FOV of the high resolution instruments observing at the DST. 

\begin{figure}
\begin{center}
\includegraphics[trim=0 120 0 0, clip, scale=0.6]{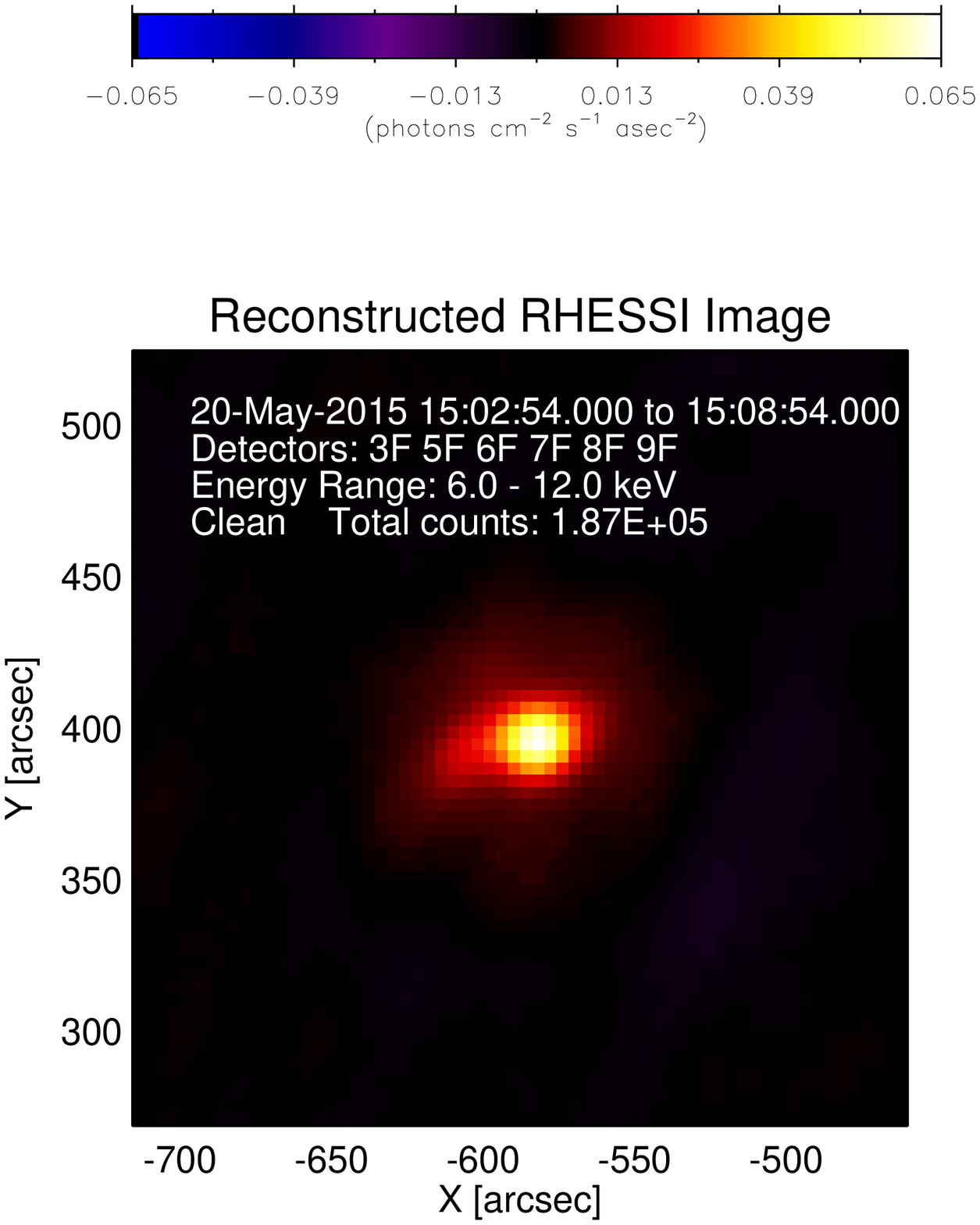}
\caption{Reconstructed image obtained by RHESSI from 15:02:54 UT to 15:08:54 UT in the energy band between 6.0 and 12.0 keV using the CLEAN algorithm.} 
\label{Fig8}
\end{center}
\end{figure}

Another important result comes from RHESSI, whose reconstructed image obtained using CLEAN algorithm \citep{Hur02} in the energy band between 6.0 and 12.0 KeV from 15:02:54 UT to 15:08:54 UT shows a low energy X-ray source located at [-582\arcsec$,$397\arcsec] (Figure \ref{Fig8}). This corresponds to the center of IBIS FOV and to the described magnetic field system. Moreover, in the RHESSI reconstructed image a brighter region departing from the main source and directed towards South-West is also visible. This feature seems to correspond to the shape, direction, and location of the outer spine outlined by the potential field extrapolation.

\section{Summary}

Despite the low intensity of the flare described in this Paper, its relevance can be ascribed to the opportunity of investigate the characteristics of the chromospheric ribbons by means of high resolution observations. In fact, ground based observations in the \ion{Ca}{2} 854.2 nm line by IBIS and in the H$\beta$ 486.1 nm line by ROSA allowed us to study the evolution of the emission in the chromosphere during the main phase of the flare. The main new results can be summarized as follows.

\begin{enumerate}[(i)]

\item We observed the brightening of a semi-circular ribbon formed by several adjacent bright kernels. The diameter of this semi-circular ribbon was of about 40\arcsec, while the time necessary to its complete brightening was of about 10 minutes, from the beginning of the flare at 15:03 UT to 15:13 UT ({\em top right panel} of Figure \ref{Fig5}).

\item The kernels had a diameter of 1\arcsec-2\arcsec{} and started to bright sequentially in clockwise direction from the East to the West. The brighter kernels were located at the outer footpoints of the darker filaments.

\item On the base of the potential field extrapolation we attributed the flare to a magnetic reconnection event occurred in a magnetic field configuration containing a 3D null point. We found a clear spatial correspondence between the circular ribbon and the fan footpoints, as well as between the other brightenings and the footpoints of the inner and outer spines. 

\item From SDO/AIA images it seems that the outer spine started to activate before the flare onset. In fact, we observed a few minutes before the beginning of the flare the brightening of a loop connecting the western part of the AR with its eastern part (see the {\em top left} panel of Figure \ref{Fig2bis}).  

\item Although we did not observe the beginning of the event by IBIS and ROSA, we observed some LOS plasma motions towards the observer in the \ion{Ca}{2} 854.2 line which can be interpreted as the last phase of the eruption of a small flux rope corresponding to the complex bundle of loops observed by AIA/SDO at 171 \AA{} during the flash phase of the flare. Instead, the main blob of plasma moving away from the observer with a velocity of about 10 km s$^{-1}$ seems to correspond to the central flare kernel, i.e., to the acceleration site of the plasma along the inner spine.

\item During the main phase of the flare RHESSI detected a significant emission between 6.0 and 12.0 keV in a region corresponding to the 3D null point and along the outer spine. This low energy X-ray emission seems to indicate that the heating source was around the null point and along the outer spine, while it did not show any spatial correlation with the chromospheric emission, contrarily to what observed by \citet{Yan15} in a X-class circular ribbon flare occurred on 2013 October 23.

\end{enumerate}

\section{Conclusions}

The event studied in this work showed several interesting signatures useful to model the magnetic reconnection in and around a 3D null point. The very high-quality of the dataset allowed us to investigate in detail the strict correspondence between the complex magnetic topology associated to a 3D null point and the observable features during the flare. 

The clockwise propagation of the semi-circular ribbon emission with a diameter of about 40\arcsec{} in about 10 minutes suggests that slipping/slip-running reconnection processes occur along the fan \citep{Aul06, Pon13}, as \citet{Mas09} and \citet{Rei12} observed during a confined flare occurred on 2002 November 16. However, the fact that the semi-circular ribbon is formed by adjacent bright kernels and that the brighter ones correspond to the outer footpoints of the darker filaments is a signature of the inhomogeneity distribution of the magnetic field forming the fan.   

Unfortunately, for our event we do not have any information about the delay of the remote brightening because it is located outside the FOVs of the high resolution instruments and AIA/SDO images do not have enough spatial and temporal resolution to detect it. However, we observed the brightening of a EUV loop corresponding to the outer spine about 10 min before the beginning of the flare. This feature suggests that this kind of event may be preceded by other resistive instabilities along the spines, although the role of these instabilities should be investigated by new models. 

Another interesting issue raised by our study concerns the filament material observed near the main ribbon. Like in the circular ribbon flares studied by \citet{Wan12}, in this event a central parasitic magnetic field is encompassed by the opposite polarity, forming a circular polarity inversion line (PIL). However, our high resolution observation in the chromospheric lines allowed us to interpret the filament material along the PIL not to be a single circular filament, but to be formed by several filaments connecting the main pore, in the center of the FOV, to the surrounding ribbons, i.e., along the same direction of the fan loops. 

\citet{Den13} already presented high resolution observations of a flare characterized by a main circular ribbon, but in that case they observed with IBIS only the remote ribbon. They found a temporal correlation between the H$\alpha$ and hard X-rays emission at 12-25 keV during the flare impulsive phase. A new finding obtained in our study is that the new models of the magnetic reconnection around a 3D null point should take into account also the emission in the low energy X-ray not only in the location of the null point but also along the outer spine.

In conclusion, we think that the low intensity of the flare might be ascribed to the low amount of magnetic flux involved in the event, although, the symmetry of the magnetic system might be also responsible of the small energy released during the flare. In fact, following \citet{Wyp16} we estimated a ratio N/L of about 0.8, where N is the diameter of the semi-circular ribbon (about 40\arcsec) and L is the distance between the footpoints of the coronal loop became brighter a few minutes before the onset of the flare (about 50\arcsec).

At present, we guess that the next generation facilities, such as the GREGOR telescope \citep{Sch12a}, the Daniel K. Inouye Solar Telescope \citep[formerly the Advanced Technology Solar Telescope][]{Kei10}, and the European Solar Telescope \citep{Col10}, will provide the opportunity to deep the observational aspects of these kind of events by means of a wider sample of case studies and a superlative spatial resolution.

\acknowledgments

The authors wish to thank the DST staff for its support during the observing campaigns. In particular, P.R. is grateful to Mike Bradford who allowed him to catch this interesting event. The research leading to these results has received funding from the European Commission's Seventh Framework Programme under the grant agreements SOLARNET (project no. 312495) and F-Chroma (project no. 606862). This work was also supported by the Istituto Nazionale di Astrofisica (PRIN-INAF-2014) and bu Universit\'{a} degli Studi di Catania (PRIN-MIUR-2012).


%% in parentheses, after the keyword, but they will not be verified.

{\it Facilities:} \facility{DST (IBIS), DST (ROSA), SDO (AIA, HMI)}.



%\appendix

%\section{Appendix material}

\clearpage


\begin{thebibliography}

\bibitem[Alissandrakis(1981)]{Ali81} Alissandrakis, C. E. 1981, A\&A, 100, 197

\bibitem[Aulanier et al.(2006)]{Aul06} Aulanier, G., Pariat, E., D´emoulin, P., \& Devore, C. R. 2006, Sol. Phys., 238, 347

\bibitem[Baumann, Galsgaard \& Nordlund(2013)]{Bau13a} Baumann, G., Galsgaard, K., \& Nordlund, \AA., 2013a, Sol. Phys., 284, 467

\bibitem[Baumann, Haugbolle \& Nordlund(2013)]{Bau13b} Baumann, G., Haugbolle, T., \& Nordlund, \AA., 2013b, \apj, 771, 93

\bibitem[Beveridge(2006)]{Bev06} Beveridge, C. 2006, Sol. Pys., 236, 41

\bibitem[Cavallini(2006)]{Cav06} Cavallini, F. 2006, Sol. Phs., 236, 415

\bibitem[Collados(2010)]{Col10} Collados, M., Bettonvil, F., Cavaller, L. \& EST Team 2010, AN, 331, 615

\bibitem[Deng et al.(2013)]{Den13} Deng, N., Tritschler, A., Jing, J. et al. 2013, \apj, 769, 112

\bibitem[Hoeksema et al.(2014)]{Hoe14} Hoeksema, J. T., Liu, Y., Hayashi, K., et al. 2014, Sol. Phys., 289, 3483

\bibitem[Hurford et al.(2002)]{Hur02} Hurford, G. J., Schmahl, E. J., Schwartz, R. A., et al. 2002, Sol. Phys., 210, 61

\bibitem[Jess et al.(2010)]{Jes10} Jess, D., Mathioudakis, M., Christian, D., et al. 2010, Sol. Phys., 261, 363

\bibitem[Jiang et al.(2014)]{Jia14} Jiang, C., Wu, S. T., Feng, X., \& Hu, Q. 2014, \apj, 780, 55

\bibitem[Keil et al.(2010)]{Kei10} Keil, S. L., Rimmele, T. R., Wagner, J. \& ATST Team 2010, AN, 331, 609

\bibitem[Galsgaard \& Nordlund(1997)]{Gal97} Galsgaard, K., \& Nordlund, Å. 1997, J. Geophys. Res., 102, 231

\bibitem[Galsgaard, Priest \& Titov(2003)]{Gal03} Galsgaard, K., Priest, E. R., \& Titov, V. S. 2003, J. Geophys. Res. (Space Physics), 108, 1042

\bibitem[Guglielmino et al.(2016)]{Gug16} Guglielmino, S. L., Zuccarello, F., Romano, P., Cristaldi, A., Ermolli, I., Criscuoli, S., Falco, M., \& Zuccarello, F. P. 2016, \apj, 819, 157

\bibitem[Liu et al.(2015)]{Liu15} Liu, C., Deng, N., Liu, R., et al. 2015, \apj, 769, 112

\bibitem[L\"ofdahl(2002)]{Lof02} L\"ofdahl, M.G. 2002, Proc. SPIE, 4792, 146L

\bibitem[Mandrini et al.(2014)]{Man14} Mandrini, C. H., Schmieder, B., Démoulin, P., Guo, Y., \& Cristiani, G. D. 2014, Sol. Phys., 289, 2041

\bibitem[Masson et al.(2009)]{Mas09} Masson, S., Pariat, E., Aulanier, G., \& Schrijver, C. J. 2009, \apj, 700, 559

\bibitem[Masson et al.(2012)]{Mas12} Masson, S., Aulanier, E., Pariat, E., \& Klein, K.L. 2012, Sol. Phys., 276, 199

\bibitem[Pariat, Antiochos \& DeVore(2009)]{Par09} Pariat, E., Antiochos, S. K., \& DeVore, C. R. 2009, \apj, 691, 61

\bibitem[Pontin, Bhattacharjee \& Galsgaard(2007)]{Pon07a} Pontin, D. I., Bhattacharjee, A., \& Galsgaard, k. 2007a, Phys. Plasmas, 14, 052106

\bibitem[Pontin \& Galsgaard(2007)]{Pon07b} Pontin, D. I., \& Galsgaard, K. 2007, J. Geophys. Res., 112, 3103

\bibitem[Pontin, Priest \& Galsgaard(2013)]{Pon13} Pontin, D. I., Priest, E., \& Galsgaard, K. 2013, \apj, 774, 154

\bibitem[Pontin, Galsgaard \& D´emoulin(2016)]{Pon16} Pontin, D. I., Galsgaard, K., \& D´emoulin, P. 2016, Sol. Phys., 291, 1739

\bibitem[Reid et al.(2012)]{Rei12} Reid, H. A. S., Vilmer, N., Aulanier, G., \& Pariat, E. 2012, A\&A, 547, A52

\bibitem[Romano et al.(2013)]{Rom13} Romano, P., Frasca, D., Guglielmino, S. L., et al. 2013, \apj, 771, L3

\bibitem[Ruiz Cobo \& del Toro Iniesta(1992)]{Rui92} Ruiz Cobo, B., \& del Toro Iniesta, J. C. 1992, \apj, 398, 375

\bibitem[Savcheva et al.(2015)]{Sav15} Savcheva, A., Pariat, E., McKillop, S., et al. 2015, \apj, 810, 96

\bibitem[Schmidt  et al.(2012)]{Sch12a} Schmidt, W., von der Luhe, O., Volkmer, R., et al. 2012, AN, 333, 796

\bibitem[Schou et al.(2012)]{Sch12} Schou, J., Scherrer, P. H., Bush, R. I., et al. 2012, Sol. Phys., 275, 229

\bibitem[Sun et al.(2013)]{Sun13} Sun, X., Hoeksema, J. T., Liu, Y., et al. 2013, \apj, 778, 139

\bibitem[Wang \& Liu(2012)]{Wan12} Wang, H., \& Liu, C. 2012, \apj, 760, 101

\bibitem[Wyper \& DeVore(2016)]{Wyp16} Wyper, P. F., \& DeVore, C. R. 2016, \apj, 820, 77

\bibitem[Yang, Guo \& Ding(2015)]{Yan15} Yang, K., Guo, Y., Ding, M.D. 2015, \apj, 806, 171

\bibitem[Zhang et al.(2015)]{Zha15} Zhang, Q. M., Ning, Z. J., Guo, Y., et al. 2015, \apj, 805, 4



\end{thebibliography}
\end{document}